\documentclass[a4paper,fleqn,usenatbib]{mnras}   

\usepackage[T1]{fontenc}
\usepackage{ae,aecompl}
\usepackage{graphicx}	
\usepackage{amsmath}	
\usepackage{amssymb}	

\newcommand{\nvalone}{\textrm{N}\textsc{v}}
\newcommand{\civalone}{\textrm{C}\textsc{iv}}

\newcommand{\hi}{\textrm{H}\textsc{i}}

\newcommand{\oii}{[\textrm{O}\textsc{ii}]}
\newcommand{\heiialone}{[\textrm{He}\textsc{ii}]}
\newcommand{\oiidoub}{[\textrm{O}\textsc{ii}]\ensuremath{\lambda3727,3729}}
\newcommand{\oiilam}{[\textrm{O}\textsc{ii}]\ensuremath{\lambda3727}}
\newcommand{\oiiiv}{[\textrm{O}\textsc{iii}]\ensuremath{\lambda5007}}

\newcommand{\oiiidoub}{[\textrm{O}~\textsc{iii}]\ensuremath{\lambda\lambda4959,5007}}
 
\newcommand{\ha}{\ifmmode {\rm H}\alpha \else H$\alpha$\fi}
\newcommand{\hb}{\ifmmode {\rm H}\beta \else H$\beta$\fi}
\newcommand{\lya}{\ifmmode {\rm Ly}\alpha \else Ly$\alpha$\fi}
\newcommand{\pg}{\ifmmode {\rm P}\gamma \else Pa$\gamma$\fi}
\newcommand{\lyb}{\ifmmode {\rm Ly}\beta \else Ly$\beta$\fi}
\newcommand{\lyg}{\ifmmode {\rm Ly}\gamma \else Ly$\gamma$\fi}

\newcommand{\siii}{[\textrm{Si}\textsc{ii}]\ensuremath{\lambda1260}}
\newcommand{\oi}{[\textrm{O}\textsc{i}]\ensuremath{\lambda1303}}
\newcommand{\cii}{[\textrm{C}\textsc{ii}]\ensuremath{\lambda1334}}
\newcommand{\sivdoub}{[\textrm{S}\textsc{iv}]\ensuremath{\lambda1393,1402}}
\newcommand{\sivblue}{[\textrm{S}\textsc{iv}]\ensuremath{\lambda1393.76}}
\newcommand{\sivred}{[\textrm{S}\textsc{iv}]\ensuremath{\lambda1402.77}}
\newcommand{\ciistar}{[\textrm{C}\textsc{ii}]\ensuremath{^*\lambda1335.71}}
\newcommand{\nv}{\textrm{N}\textsc{v}\ensuremath{\lambda1240}}
\newcommand{\nvdoub}{[\textrm{N}\textsc{V}]\ensuremath{\lambda\lambda1239,1243}}

\newcommand{\heii}{\textrm{He}\textsc{ii}\ensuremath{\lambda1640}}

\newcommand{\flyc}{\ifmmode  \mathrm{f}_\mathrm{esc}\mathrm{(LyC)} \else $\mathrm{f}_\mathrm{esc}\mathrm{(LyC)}$\fi}

\def\kms{km s$^{-1}$}

\def\erg{erg s$^{-1}$ cm$^{-2}$ \AA$^{-1}$}
\def\ergs{\ifmmode \mathrm{erg\hspace{1mm}s}^{-1} \else erg s$^{-1}$\fi}
\def\ergscm{erg s$^{-1}$ cm$^{-2}$}
\def\micron{\ifmmode \mu\mathrm{m} \else $\mu$m\fi}
\def\msun{\ifmmode \mathrm{M}_{\odot} \else M$_{\odot}$\fi}
\def\msunyr{\ifmmode \mathrm{M}_{\odot} \hspace{1mm}{\rm yr}^{-1} \else $\mathrm{M}_{\odot}$ yr$^{-1}$\fi}
\def\zsun{\ifmmode Z_{\odot} \else Z$_{\odot}$\fi}
\def\lsun{\ifmmode L_{\odot} \else L$_{\odot}$\fi}
\def\mstar{\ifmmode \mathrm{M}_{\star} \else M$_{\star}$\fi}

\title[The highest redshift stellar ioniser]
{Direct Lyman continuum and Lyman $\alpha$ escape observed at redshift 4}

\author [E.~Vanzella et al.]{
\parbox[t]{\textwidth}{E.~Vanzella$^1$\thanks{E-mail: eros.vanzella@oabo.inaf.it},
M.~Nonino$^2$, G.~Cupani$^2$, M.~Castellano$^3$, E.~Sani$^4$, M.~Mignoli$^1$, F.~Calura$^1$, M.~Meneghetti$^1$, R.~Gilli$^1$, A.~Comastri$^1$, A.~Mercurio$^5$, G.~B.~Caminha$^6$, K.~Caputi$^6$, P.~Rosati$^{1,7}$, C.~Grillo$^{8,9}$, S.~Cristiani$^{2}$, I.~Balestra$^{10}$, A.~Fontana$^3$, and M.~Giavalisco$^{11}$
}
\vspace*{8pt}\\
$^1$INAF -- Osservatorio Astronomico di Bologna, via Gobetti 93/3, 40129 Bologna, Italy\\
$^2$INAF -- Osservatorio Astronomico di Trieste, via G. B. Tiepolo 11, I-34143, Trieste, Italy\\
$^3$INAF -- Osservatorio Astronomico di Roma, Via Frascati 33, I-00078 Monte Porzio Catone (RM), Italy\\
$^4$European Southern Observatory, Alonso de Cordova 3107, Casilla 19, Santiago 19001, Chile \\
$^5$INAF -- Osservatorio Astronomico di Capodimonte, Via Moiariello 16, I-80131 Napoli, Italy\\
$^6$Kapteyn Astronomical Institute, University of Groningen, Postbus 800, 9700 AV Groningen, The Netherlands\\
$^7$Dipartimento di Fisica e Scienze della Terra, Universit\`a degli Studi di Ferrara, via Saragat 1, I-44122 Ferrara, Italy\\
$^8$Dipartimento di Fisica, Universit\`a  degli Studi di Milano, via Celoria 16, I-20133 Milano, Italy\\
$^9$Dark Cosmology Centre, Niels Bohr Institute, University of Copenhagen, Juliane Maries Vej 30, 2100 Copenhagen, Denmark\\
$^{10}$University Observatory Munich, Scheinerstrasse 1, D-81679 Munich, Germany\\
$^{11}$Astronomy Department, University of Massachusetts, Amherst, MA 01003, USA\\
}

\begin{document}
\date{}
\maketitle

\begin{abstract}
 We report on the serendipitous discovery of a $z=4.0$, $M_{1500}=-22.20$ star-forming galaxy ({\em Ion3}) showing copious 
 Lyman continuum (LyC) leakage ($\sim 60$\% escaping), a remarkable multiple peaked \lya\ emission, and
 significant \lya\ radiation directly emerging at the resonance frequency. 
 This is the highest redshift confirmed LyC emitter in which the ionising and \lya\ radiation possibly share a common 
 ionised channel (with $N_{HI}<10^{17.2}$ cm$^{-2}$).
 {\em Ion3} is spatially resolved, it shows clear stellar winds signatures like the P-Cygni \nv\ profile,
 and has blue ultraviolet continuum ($\beta = -2.5 \pm 0.25$, $F_{\lambda} \sim \lambda^{\beta}$) with weak
 low-ionisation interstellar metal lines.
 Deep VLT/HAWKI Ks and Spitzer/IRAC 3.6$\mu m$ and 4.5$\mu m$ 
 imaging show a clear photometric signature of the \ha\ line with equivalent width of 1000\AA~rest-frame
 emerging over a flat continuum (Ks$-4.5\mu m \simeq 0$). 
 From the SED fitting we derive a stellar mass of  $1.5\times 10^{9}$\msun, SFR of 140 \msunyr\ and
 age of $\sim 10$ Myr, with a low dust extinction, E(B-V)$\lesssim 0.1$, placing the source in the
 starburst region of the SFR$-$M$^{*}$ plane. 
 {\em Ion3} shows similar properties of another LyC emitter previously 
 discovered ($z=3.21$, {\em Ion2}, Vanzella et al. 2016). 
 {\em Ion3} (and {\em Ion2}) represents ideal high-redshift reference cases to guide
 the search for reionising sources at $z>6.5$ with JWST. 
\end{abstract}


\section{Introduction}
The definition of a reference sample of Lyman continuum (LyC) emitters at $z\lesssim4.5$ 
is crucial to guide the identification of the sources that reionised the universe 
at $z>6.5$, an epoch when the LyC is not directly observable \citep{worseck14}.
Several LyC leakers showing escape fraction of $f_{esc}^{abs} \sim 4-15\%$ have been confirmed  
 in the nearby universe, $z \sim0-0.3$  (see \citealt{izotov16a,izotov16b} and references therein).
Recently \citet{izotov17} identified another LyC emitter with $f_{esc}\simeq 46\%$, the highest value currently
measured for the local sample. Most searches at high redshift have proved unsuccessful, due to contamination by
foreground sources \citep{siana15}, or have yielded fairly stringent limits of $f_{esc}<0.1$ 
\citep[e.g.,][]{grazian17,vanz12}.
However, three LyC emitters have been confirmed at cosmological distances,
$2.5<z<3.2$ with $f_{esc} \sim 30- 100\%$ \citep{vanz16, debarros16,shapley16,bian17}.
As shown in \citet{verhamme17} the available sample of LyC leakers, both at low and high redshift,
show very consistent observational features. Such features include the high \lya\ equivalent width EW~$>40$\AA, a high
ratio of \oiiiv / \oiilam, and narrow signatures in the \lya\ emission like narrow double peaked profiles ($< 300-400$\kms) 
or \lya\ emission close to the systemic velocity ($<100$\kms). Also intense nebular optical lines  
\oiiidoub + \hb\ with EW of 1000\AA$-$1500\AA~rest-frame seem often associated to LyC emitters
\citep[e.g.,][]{schaerer16}, as well as their compact star-forming region \citep{h11} 
and the weakness of the low-ionisation interstellar absorption lines \citep[e.g.,][]{jones13, chisholm17}.  
Overall, these quantities correlate with $f_{esc}$, as was predicted by radiation transfer
models \citep{verhamme15} and photoionisation models
\citep[e.g.,][]{jaskot13, nakajima14, zackrisson13}. 
In this work we report on a serendipitously discovered LyC emitter at redshift 4, dubbed here {\em Ion3},
the highest redshift case currently known.
We assume a flat cosmology with $\Omega_{M}$= 0.3,
$\Omega_{\Lambda}$= 0.7 and $H_{0} = 70$ km s$^{-1}$ Mpc$^{-1}$.

 \begin{figure}
\centering
\includegraphics[width=8.5cm]{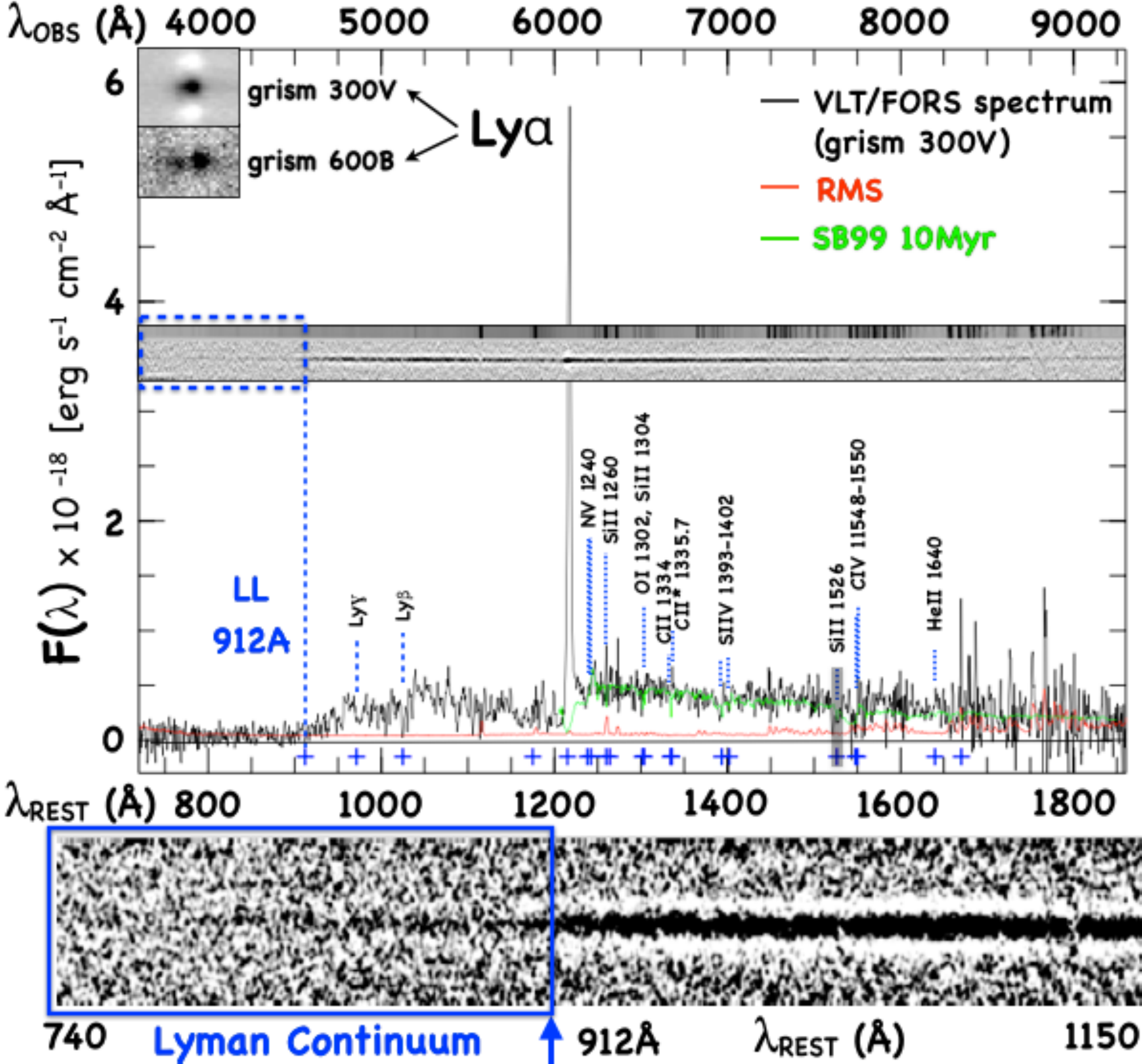}
\caption{The FORS spectrum at resolution $dv \simeq 580$ \kms\ (grism 300V) of {\em Ion3} is shown
along with the most relevant ultraviolet lines. In the top-left insets the \lya\ spectra at resolution $dv \simeq$ 300(580) \kms\ 
obtained with the grism 600B(300V) are shown. The two-dimensional signal to noise and sky spectra are shown
in the middle of the figure, in which the wavelength coverage up to 9300\AA~is well sampled, despite the fringing
pattern at $\lambda>7800$\AA~typical of the E2V blue-optimised CCD. In the bottom the 
the emerging LyC with $S/N>10$ is clearly detected.}
\label{fig1}
\end{figure}

\section{Data: FORS and X-Shooter observations}

{\em Ion3} was discovered during a FORS2 spectroscopic program executed in visitor mode during
the period $15-19$ September 2017 (prog. 098.A-0804(B), P.I. Vanzella). 
{\em Ion3} is a relatively bright object and was inserted as a filler in the MXU FORS mask 
(I-band magnitude $23.64\pm0.38$) and located at $3'41''$ from the Frontier Field galaxy cluster
AS1063 \citep{lotz17}, which lies outside the HST coverage. 
Given the large separation from the galaxy cluster the resulting magnification is low,
a well constrained $\mu = 1.15 \pm 0.02$ (or $\Delta m = 0.15$, \citealt{caminha16}). 
In the following all the reported quantities are corrected for $\mu$.

The data reduction was carried on as described in several previous works 
\citep[e.g.,][]{vanz14} in which the AB-BA sky subtraction scheme was
implemented. The final spectrum consists of 14 hours integration with an average seeing of 
$0.8''$ and spectral resolution $dv \simeq 600$ \kms\ at $\lambda=7000$\AA~(R=$\lambda/d\lambda$=500, grism 300V). 
Figure~\ref{fig1} shows the FORS2 spectrum covering the wavelength range 3700\AA$-$9300\AA.
During the same program 098.A-0804(B) an additional one hour integration was obtained with
the grism 600B, doubling the spectral resolution ($dv \simeq 300$ \kms, R=1000). 
We clearly confirmed the double peaked \lya\ profile of {\em Ion3} not well resolved with the
300V grating  (see Figure~\ref{fig1}, top-left).
Additional four hours X-Shooter integration on {\em Ion3} was subsequently obtained
during November 2017 with an average seeing conditions of $0.8''$ (prog. 098.A-0665, P.I. Vanzella), 
providing a final spectrum spanning the range 3400\AA~-~24000\AA~with spectral resolution $dv \simeq 35-60$\kms. 
We refer the reader to \citet{vanz14, vanz17} for details about FORS and X-Shooter data reduction. 

\section{Results}

\subsection{The Lyman continuum emission}

The most intriguing feature emerging from {\em Ion3} is the LyC leakage 
at $\lambda<912$\AA~rest-frame detected in the FORS spectrum with S/N=6.4(11.1) if averaged over the interval $880-910$($800-910$\AA), 
and corresponding to magnitude 27.5 (AB).  Figure~\ref{fig1} shows the two-dimensional spectrum with the 
LyC signal spatially aligned with the non-ionising radiation. The possible presence of a 
foreground object at $z<4$ mimicking the LyC
signal represents a serious problem in this kind of observations \citep[e.g.,][]{siana15, vanz12}, especially
when HST imaging is not available. However we can reasonably exclude such an interloper.
First, we note that the probability chance of a superposition is low, less than 1\%  adopting the above magnitude and the
seeing of $0.8''$ (see \citealt{vanz10}). Second, there is no trace of any spectral line arising
from a foreground object, both in the deep FORS spectrum (that excludes \oiidoub\ at $z<1.45$ and \lya\ at $z>1.9$)
and wide X-Shooter spectrum, that easily would have captured several UV and/or optical rest-frame emission lines
in the redshift range $0<z<4$. 
Moreover,  all the properties of {\em Ion3} support a very low column density of neutral gas along the
line of sight, making the entire picture consistent with the emerging LyC signal. 

From the FORS spectrum the observed fluxes at 900\AA~and 1500\AA~rest-frame are 
$F_{\lambda}^{900} = (5.6 \pm 0.08) \times 10^{-20}$\erg~and $F_{\lambda}^{1500} = (3.8 \pm 0.09)\times 10^{-19}$\erg, respectively,
corresponding to a flux density ratio of $f_{\nu}(1500) / f_{\nu}(900) = 19.1\pm3.3$. 
Following \citet{vanz12} this ratio translates to a relative escape fraction {\em fesc,rel} = 20-100\%, assuming an
intrinsic ratio of the luminosity densities $L_{\nu}(1500)/L_{\nu}(900) = 1-5$ and the median
intergalactic medium (IGM) transmission at 900\AA, $<T_{IGM}^{900}>=0.26$ (with 
a central 68\% interval of $0.05-0.40$, \citet{inoue14,vanz15}.
Being a single line of sight the real $T_{IGM}^{900}$ is unknown, 
therefore, any combination of $L_{\nu}(1500)/L_{\nu}(900)>1$ and IGM transmission lower than 100\% produces
{\em fesc,rel} in the range 10\% $-$ 100\%, with a fiducial value of 60\% adopting median
 $<T_{IGM}^{900}>=0.26$ and $L_{\nu}(1500)/L_{\nu}(900) = 3$.
The inferred ionising photon production rate from $F_{\lambda}^{900}$ is $N_{phot}(900)= 3.5\times 10^{53} s^{-1}$, 
which compared to Starburst99 models for instantaneous bursts 
(Salpeter IMF, \citet{salpeter55}, and upper mass limit of $100 M_{\odot}$, \citealt{leitherer14}) yields a stellar
mass involved in the starburst event  of 
$4\times 10^{6} M_{\odot}$ with the age not larger than 20Myrs, and a number of O-type stars 
dominating the ionising radiation of  $\simeq 1.6\times 10^{4}$ (with uncertainties mainly
dominated by the aforementioned IGM transmission). 

\begin{figure}
\centering
\includegraphics[width=8.5cm]{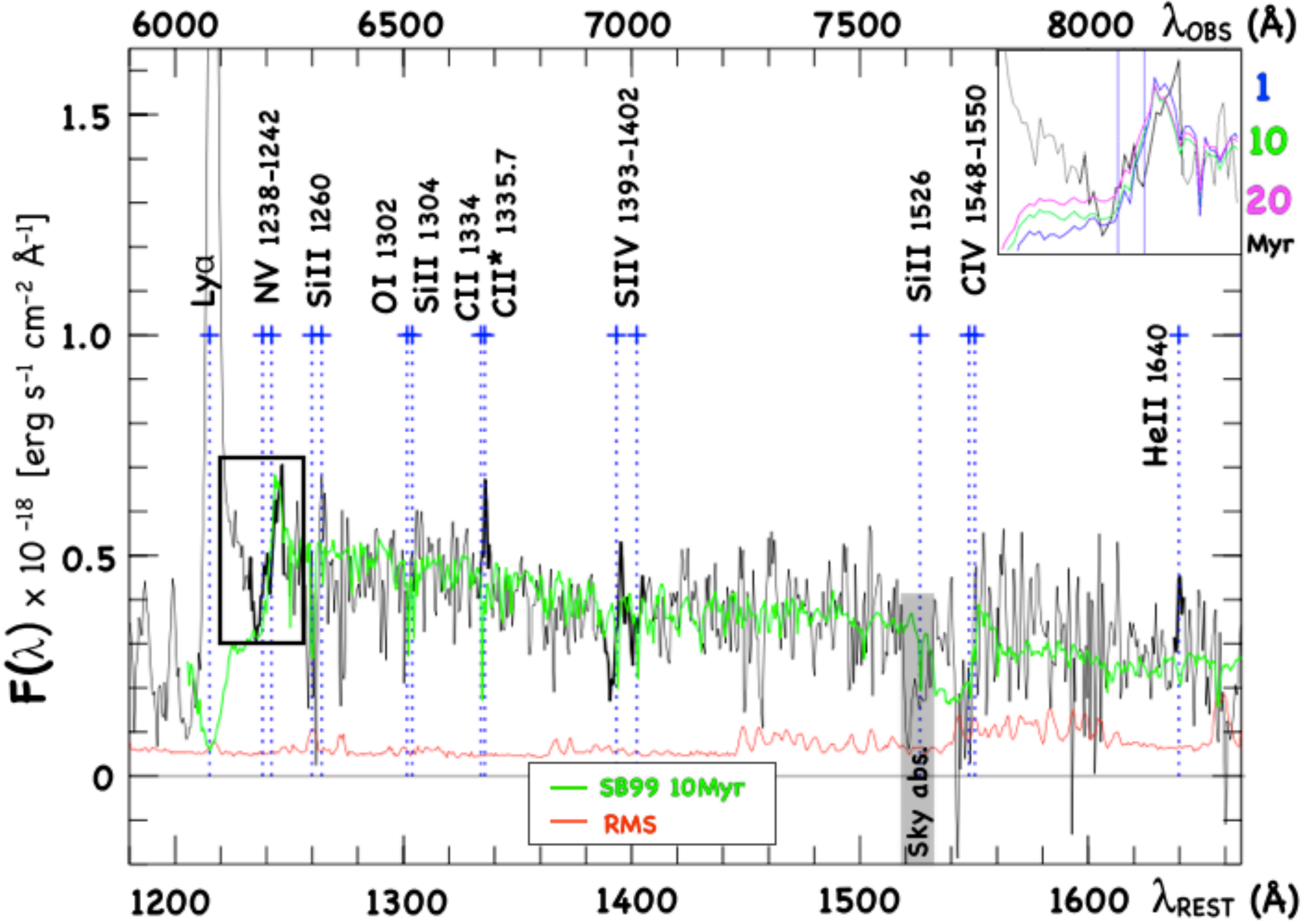}
\caption{The FORS spectrum is shown with the main spectral features reported at the systemic redshift
(blue crosses). In particular the P-Cygni  \nv, \ciistar, \sivdoub, and the \heii\ lines are highlighted with thick lines. 
The \nv\ region (square) is zoomed and shown in the top-right inset (black line), in which three Starburst99 stellar models of
1, 10 and 20 Myr are superimposed to the spectrum (coloured lines).}
\label{fig2}
\end{figure}

\subsection{The Non-ionising properties}

 The  continuum redward the \lya\ line up to $\simeq 1850$\AA~rest-frame is well detected in the
 FORS spectrum (Figure~\ref{fig1} and \ref{fig2}) with a best-fit
 ultraviolet spectral slope of $\beta = -2.50$ with 68\% central interval [-2.73,-2.28] ($F_{\lambda} \sim \lambda^{\beta}$).
 Such a steep slope is not significantly affected by the atmospheric dispersion
 (the atmospheric dispersion compensator, LADC, is part of the system at UT1),
 and is fully consistent with being powered by massive and hot stars
 (O and early B stars), which are also responsible for the ionising photons. 
 The strongest spectroscopic
 signature of these stars is provided by the broad P-Cygni profiles in the resonance transitions of highly ionised
 species that arise in the stellar winds \cite[e.g.,][]{leitherer14}.
 In our FORS spectrum a \nv\ P-Cygni profile is clearly present (see Figure~\ref{fig2}),
 observed also in local star-forming galaxies (e.g., \citealt{h11}, see also Figure~4 of \citealt{jaskot17}).
 In particular, the height and depth of the \nv\ line suggest an age of the burst
 of a few Myr. 
 We superimposed a Starburst99 \citep{leitherer14} spectral population synthesis
 model to the FORS spectrum with instantaneous burst with age 1, 10 and 20 Myr old, 0.4 solar metallicity
 (Salpeter IMF and 100$M_{\odot}$ upper mass limit), which reproduces the observations well 
 (especially the 1-10 Myr old templates, Figure~\ref{fig2}).
 Low ionisation interstellar absorption lines like \siii, \oi\ and \cii\ are very weak or even absent,
 indicating a very low gas covering fraction consistent with the measured LyC leakage \citep{h11,jones13,chisholm17}.
 Interestingly, {\em Ion3} also shows the non-resonant fluorescent emission line \ciistar\  (see Figure~\ref{fig2}).  
 Such a feature has been detected by \citet{jaskot14} on local LyC candidates
 and interpreted as an evidence of the complex geometry of the neutral gas outside the line of sight,
 like anisotropic ionising emission. 
 The systemic redshift $z_{sys}$ has been derived from the \ciistar\  line and more accurately from the X-Shooter 
 detection of the \oiidoub\ doublet, providing $z_{sys}=3.999 \pm 0.001$.
 At such redshift the doublet \nvdoub\ lies in the middle of its P-Cygni profile and the possible \heii\ line is also recognised, 
 though with low significance, 2.7$\sigma$.
 Table~\ref{tab} summarises the most relevant spectroscopic properties. 

\begin{figure}
\centering
\includegraphics[width=8.5cm]{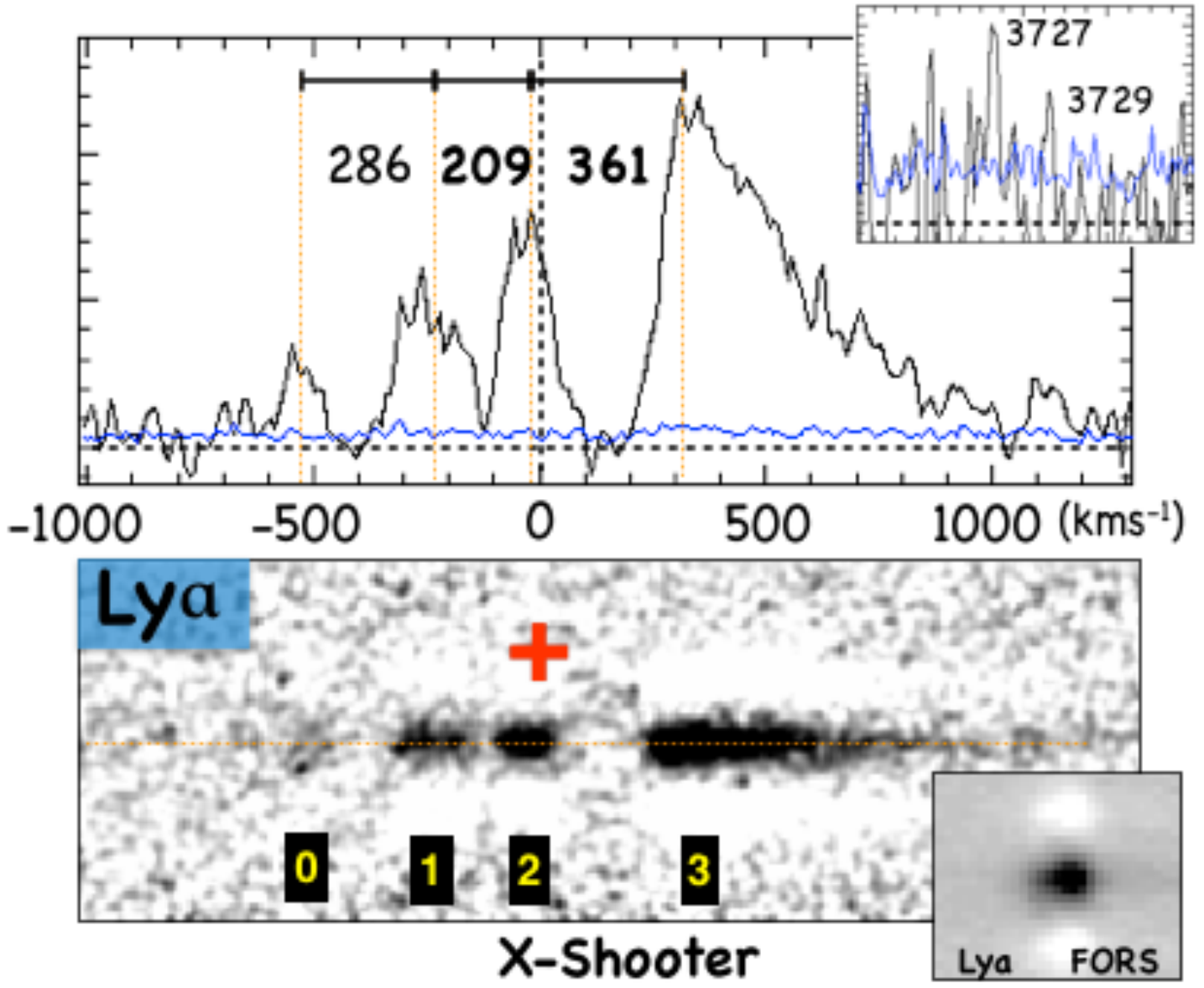}
\caption{{\it Top:} The one-dimensional X-Shooter spectrum of the \lya\  is shown (at $dv \simeq 35$ \kms) in the velocity
space (with reported the relative velocities among the peaks, \kms). 
The systemic velocity is inferred from the \oiidoub\ doublet (top-right inset) and marked with the red cross. 
{\it Bottom:}  The X-Shooter two-dimensional spectrum 
is shown with the four main structures identified. In the bottom-right inset the same \lya\ line is shown 
at resolution $dv \simeq 600$ \kms\ as observed with FORS.}
\label{lya}
\end{figure}

\subsection{The multiple peaked \lya\ emission}

In the conventional scenario for \lya\ emission, \lya\ scatters many times before escaping, which significantly alters
and broadens the original line profile. The kinematics, the column density and geometry of the \hi\ gas  are
the main ingredients that shape the \lya\ emission (not to mention the dust attenuation). 
A prominent \lya\ emission line with rest-frame EW$= 40 \pm 3$\AA~is clearly detected in the FORS spectrum (Figure~\ref{fig1}). 
Figure~\ref{lya} shows the same line at the X-Shooter spectral resolution ($dv=35$ \kms),
in which we identify four emitting structures marked as 0, 1, 2 and 3. 
While the presence of blue peaks typically suggest a low column of \hi\ gas
\citep[e.g.,][]{henry15,yang16}, the emission at peak (2) is remarkable and emerges at the resonance frequency
($z=3.9984$, i.e., $ \lesssim40$\kms\ from the systemic) where the opacity to \lya\ photons would be the
highest (Figure~\ref{lya}).
This is fully consistent with a scenario in which the LyC and (part of) the \lya\ photons 
are escaping along the same optically thin direction (to LyC, $N_{HI} <10^{17.2}$ cm$^{-2}$) 
and likely from the same cavity \citep[e.g.,][]{behrens14, verhamme15, zackrisson13}.
The LyC$-$\lya\ escape through a ionised channel discussed in \citet{behrens14}  
consider the possibility that the gas is outflowing perpendicular to a galactic disk
(and is reminiscent of a wind breaking out of a galactic disk).
In this scenario the quadruply peaked \lya\ emission observed in {\em Ion3} might be
associated to a face-on disk. 
{\em Ion3} represents the highest redshift empirical evidence of such a LyC$-$\lya\ escaping mode.
The \lya\ escape degenerates with the \hi\ column and the
outflow velocity such that fast winds can mimic low columns \citep{verhamme15}.
In this case both the LyC emission and a relatively fast wind are detected. 
The evidence of an outflowing gas is imprinted in the blueshifted interstellar metal
lines \sivblue\ and \sivred, with an average $dv \simeq -400 \pm 150$\kms.
This is also consistent with the well developed red tail of peak (3)  
possibly suggesting backscattering from the receding gas.

Interestingly, a very similar direct \lya\ escape has recently been identified in a lensed $z=2.37$ galaxy by
\citet{rivera17}, in which the central peak is also at the systemic redshift and indicative of
a possible perforated channel of very low \hi\ optical depth. 

\begin{table}
\footnotesize
\caption{Spectral and physical properties of {\em Ion3}. Line fluxes are reported in units of 
$10^{-17}$ \ergscm\ (no slit losses are considered). Quantities are corrected
for the lensing magnification $\mu=1.15$. $\sigma_z$ is the redshift error on the last digit.}
\begin{tabular}{l l r} 
Line/$\lambda_{vacuum}$ & Flux($\frac{S}{N}$) & z[$\sigma_z$], Resolution($\frac{km}{s}$) \\
\hline
\lya(0) 1215.7           &  0.37(5)   & 3.9902[4], 35 (XSHO)\\ 
\lya(1) 1215.7           &  1.16(12) & 3.9950[3], 35 (XSHO)\\  
\lya(2) 1215.7           &  1.37(15) & 3.9984[3], 35 (XSHO)\\  
\lya(3) 1215.7           &  6.23(32) & 4.0033[3], 35 (XSHO)\\  
\lya(total)                  &  9.13(73) &  $-$ , 580 (FORS)\\           
\ciistar\                     &   0.14(4) & 4.000[4], 580  (FORS)\\     
\heiialone\ $\lambda 1640.42$  & 0.12(2.7)& 4.000[4], 580 (FORS)\\ 
\oii\ $\lambda 3727.09$     &   0.38(4.5) & 3.999[1], 55 (XSHO)\\ 
\oii\ $\lambda 3729.88$     &   0.25(3) & 3.999[1], 55 (XSHO)\\ 
\hline
SED-fitting output    & value &  uncertainty\\
\hline
M(stellar) [$\times 10{^9}~M_{\odot}$]  &  1.5 & [$1.4-2.2$]   \\     
Age [Myr]                             &  11 & [$10-20$]        \\    
SFR [$M_{\odot} yr^{-1}$]  &  140 & [110-150]   \\   
E(B-V)                                & $\simeq 0.1$ & [$0-0.1$] \\   
$M_{UV}$(1500) & $-22.20$  &   $\pm 0.15$\\   
\ha\ EW(rest) [\AA] &  $\simeq 1000$  & [$700-1300$]\\
\hline
Nphot(900\AA)[$s^{-1}$]     & $3.0\times 10^{53}$  & [$10^{53-54}$]\\
log$_{10}(\xi_{ion}$[Hz~erg$^{-1}$]) & 25.6 &  $25.4-25.8$ \\
$f_{esc,rel}$                        &  0.60 & [$0.10-1.00$]\\
\hline
\hline 
\end{tabular}
\label{tab}
\end{table}

\subsection{Physical properties from the SED fitting and the nature of the ionising radiation}

\begin{figure}
\centering
\includegraphics[width=8.5cm]{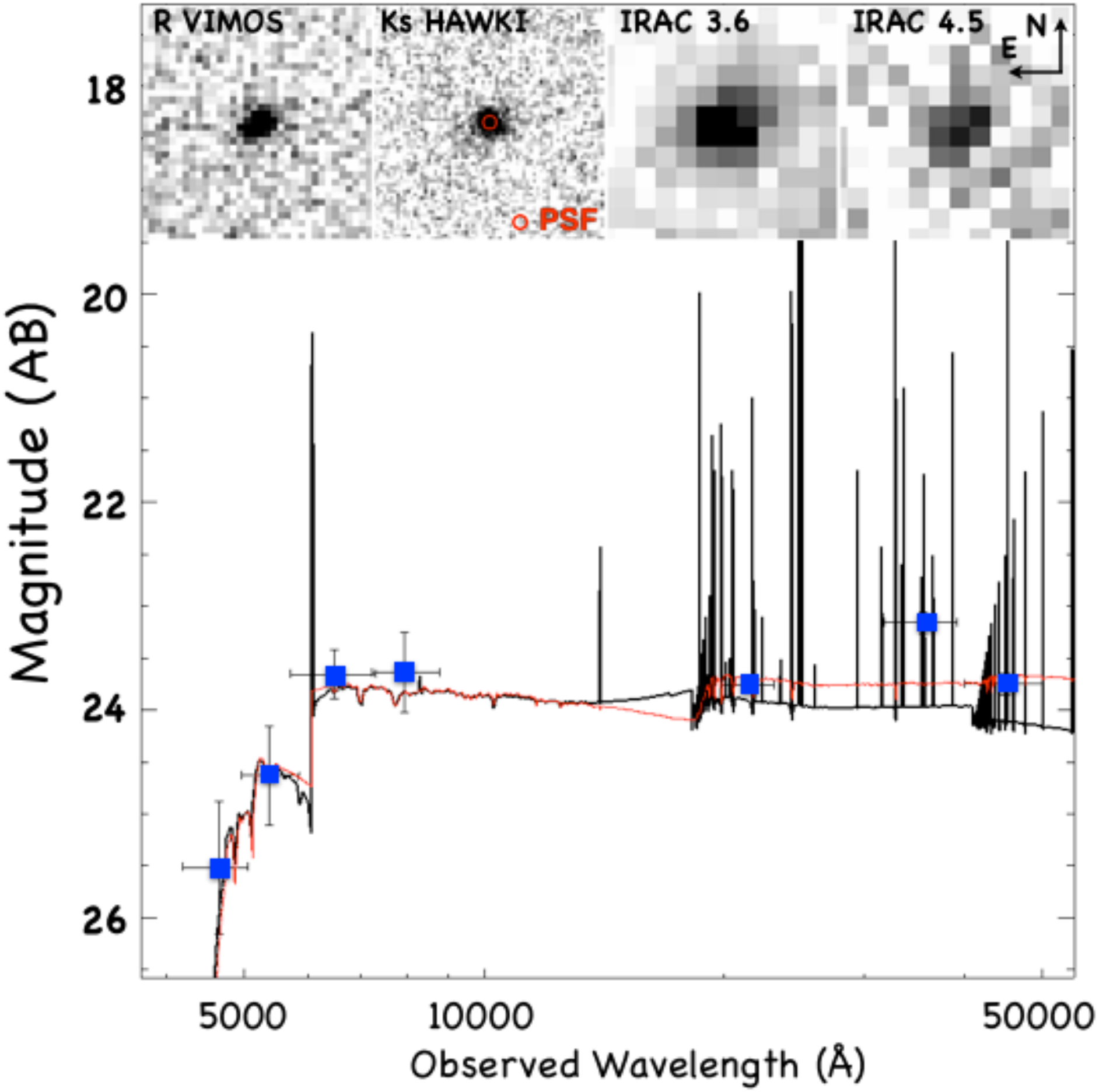}
\caption{The cutouts of {\em Ion3} (top, $6.6''$ across) and the best SED fit with only stellar (red) and 
stellar and nebular (black) templates are shown. The photometric jump at 3.6$\mu m$ is evident.}
\label{SED}
\end{figure}

SED fitting has been performed using BC03 templates \citep{BC03} 
including the nebular prescription and assuming exponentially
declining  star  formation  histories  with $e$-folding time $0.1<\tau<15$ Gyr, (see \citealt{castellano16} for details).
It has been applied to the ground$-$based ESO/WFI imaging (B$_{[842]}$, V$_{[843]}$, R$_{[844]}$ and Ic$_{[879]}$,
and deep ESO VLT/HAWKI $K_s$ (obtained with $0.39''$ seeing, \citealt{brammer16})  
and space$-$based Sptizer/IRAC 3.6$\mu m$ and 4.5$\mu m$ bands\footnote[1]{{\it http://www.stsci.edu/hst/campaigns/frontier-fields/FF-Data}}
(see Figure~\ref{SED}). While {\em Ion3} is detected at S/N$\lesssim 5$ in the ESO/WFI bands,
the optical rest-frame continuum is detected with S/N$\gtrsim 10$ in the Ks and IRAC bands. 
The most interesting features are the 
flat continuum at rest-frame wavelengths 4400\AA~and 9000\AA, 
and the clear excess in the $3.6\mu m$ band consistent with an \ha\ emission
with rest-frame EW of 1000\AA. 
The best-fit solution implies {\em Ion3} is a  relatively low stellar mass ($1.5\times 10^{9} M_{\odot}$) system undergoing
a starburst phase (SFR $\simeq 140 M_{\odot} yr^{-1}$)
consistently with the presence of prominent \lya, \nv\ P-Cygni profile,
strong \ha\ and measured LyC.
{\em Ion3} appears as a still rapidly growing system with a specific star formation
rate of $\simeq 90$Gyr$^{-1}$ (see Table~\ref{tab} for a summary of the properties of {\em Ion3}). 
The photometric estimate of the \ha\ line luminosity ($\simeq 2\times10^{43}$\ergs)
implies a high LyC photon production efficiency, $\xi_{ion}=25.6\pm0.2$[Hz~erg$^{-1}$]. 
It resembles the values derived by \citet{bouwens16} for the bluest galaxies 
($\beta < -2.3$, see also \citealt{shivaei17}), 
and consistent with the values reported from local candidate and confirmed LyC emitters,
that also show large rest-frame EW(\ha) $\sim 1000$\AA~\citep[e.g.,][]{izotov17,jaskot17}, 
and eventually similar to those inferred at $z>6$ reported by \citet{stark17}.
It is also worth noting that {\em Ion3} belongs to the starburst region of the $z\sim4.5$ 
SFR$-$M$^{*}$ bimodal distribution recently identified by \citet{caputi17}.
Based on current data, there is no evidence of, nor any need 
for, any contribution to the UV emission by an AGN. 
The relatively large ratios of \lya / \nvalone\ ($\simeq 17\pm2$) and \lya / \civalone\ $ \gtrsim 20$ tend to exclude the presence of 
an obscured AGN \citep[e.g.,][]{alex13}.
A relatively shallow Chandra exposure of 130 ksec in the 
$0.5-2$ kev band available in the field\footnotemark[1]  yields a limit of $1.6 \times 10^{-15}$ \ergscm\  ($3\sigma$),
corresponding to an upper limit to the X-ray luminosity of $3 \times 10^{43}$ erg~s$^{-1}$, 
which rules out a luminous AGN. 
Assuming a low-luminosity AGN is present, the very young burst detected
would imply the ionising photons are a mixture of stellar
and non-stellar radiation escaping along a transparent medium ($N_{HI}<10^{17.2}$ cm$^{-2}$), 
that, however, should not be able to attenuate the expected high-ionisation emission lines, making {\em Ion3} either 
a very special case among the AGN category or a pure star-forming dominated object. 
Another possibility is that
{\em Ion3} has been captured  just after the AGN has turned off, such that the optically thin channel produced by
the previous nuclear activity enables the \lya\ and LyC stellar radiation to escape toward the observer. 

\section{Final remarks}

{\em Ion3} is a bright star-forming galaxy showing copious LyC leakage identified just
1.5~Gyr after the Big-Bang ($z=4$). This makes {\em Ion3} the highest redshift confirmed LyC emitter 
known so far. In particular, 

\begin{itemize}
{\item The spectral features and the SED-fitting suggest that {\em Ion3} is a young, low mass system
undergoing a starburst phase containing hot and massive stars, 
with a specific star formation rate of 90 Gyr$^{-1}$.} 
{\item The FORS and X-Shooter spectra reveal for the first time at such redshift
a transparent ionised channel through which the \lya\ photons escape at the resonance frequency, plausibly along
the same path of the LyC photons.}

\end{itemize}

The absence of HST imaging prevents us from
deriving any conclusion on the morphology of the galaxy. 
The image with the best seeing is the VLT/HAWKI Ks-band (0.39$''$) from which {\em Ion3}
appears marginally resolved with a FWHM $\sim 2-3$ kpc proper. 
We refer the reader to a future work focused on the detailed analysis of the spatial distribution of
the ionising and non-ionising radiation, as well as a proper radiative transfer modelling of the \lya\ emission.
However, it is worth noting that {\em Ion3} shows similar properties of another
LyC emitter we identified at $z=3.21$ (dubbed {\em Ion2}, \citealt{vanz16}), e.g.,
a structured \lya\ shape, the blue UV slope and weak low-ionisation interstellar absorption lines.
While in the case of {\em Ion2} we confirmed also very strong \oiiiv\ lines (EW$>1000$\AA~rest-frame) 
and large O32 index (\oiiiv / \oiilam$>$10) making it among of the highest redshift Green Pea galaxy 
and suggesting a density-bounded condition,  {\em Ion3} might also have a perforated medium and
will need JWST to probe the rest-frame optical
wavelengths (imaging and spectroscopy) and HST to image directly the LyC.
Irrespective of the nature of the ionising radiation, {\em Ion3} represents a unique
high-redshift laboratory where ionised channels carved in the interstellar medium by one or more
feedback sources can be studied.
 {\em Ion3} and {\em Ion2} represent ideal reference cases to guide the search for reionising sources at
 $z>6.5$ with JWST. 

\section*{Acknowledgments}
We thank the referee for constructive comments. 
EV gratefully acknowledge the excellent support by ESO staff at Paranal 
during the observations. We thank G. Zamorani, A. Jaskot, S. Oey, D. Schaerer and A. Grazian for useful discussions.
F.C., A.M acknowledge funding from the INAF PRIN-SKA 2017 program 1.05.01.88.04.
MM acknowledges support from the Italian  Ministry of Foreign Affairs and
International Cooperation, Directorate General for Country Promotion.
Based on observations collected at the European Southern Observatory for Astronomical
research in the Southern Hemisphere under ESO programmes P098.A-0804(B), P098.A-0665(B).

\end{document}